\documentclass[preprint,showkeys,amsmath,amssymb]{revtex4}

\usepackage[american]{babel}
\usepackage[dvips]{graphicx}
\usepackage{bm}
\usepackage{amssymb, amsmath}
\usepackage{color}
\usepackage{dcolumn}
\DeclareSymbolFont{AMSb}{U}{msb}{m}{n}
\DeclareMathSymbol{\R}{\mathbin}{AMSb}{"52}

\usepackage{natbib} 
\usepackage{mathtools} 
\usepackage{booktabs} 

\usepackage{graphicx}  
\usepackage{dcolumn}   
\usepackage{bm}        

\usepackage{amssymb, amsmath}
\usepackage{mathrsfs}			
\usepackage{color}

 \newcommand{\ve}[1]{{\bf #1}}
 \newcommand{\vve}[1]{{\bf #1}}

\newcommand{\D}{\partial}

\newcommand{\delt}{{\Delta t}}

\newcommand{\Di}[1]{\frac {\D #1} {\D x_i}}

\newcommand{\DiDi}[1]{\frac {\D^2 #1 }{\D x_i^2 }}

\newcommand{\excl}[1]{{\backslash \hspace{-0.3em} #1}}

\newcommand{\crossout}[1]{{\slash \hspace{-0.5em} #1}}

\newcommand{\bracket}[1]{\left[#1\right]}

\newcommand{\parenth}[1]{\left(#1\right)}

\DeclareSymbolFont{AMSb}{U}{msb}{m}{n}
\DeclareMathSymbol{\R}{\mathbin}{AMSb}{"52}

\newtheorem{thm}{Theorem} 
\newtheorem{corollary}{Corollary} 

\begin{document}


\vskip 1cm

\title{An overview of the quantitative causality analysis and causal graph reconstruction based on a rigorous formalism of information flow}

\author{X.~San~Liang}
\email{sanliang@courant.nyu.edu}
\affiliation{Fudan University, Shanghai 200438, China}
\affiliation{Shanghai Qi Zhi Institute \\
(Andrew C. Yao Institute for Artificial Intelligence), \\
Shanghai 200232, China}

	\date{September 24, 2021}

\begin{abstract}
Inference of causal relations from data now has become an important
field in artificial intelligence. 
During the past 16 years, causality analysis has been developed
independently in physics from first principles, and, 
moreover, in a quantitative sense. This short note presents a brief
summary of this line of work, including part of the theory and several
representative applications.
\end{abstract}


\begin{center}
{\small Presented at the 1st Int'l AIxIA Workshop on Causality, Causal-ITALY}\\
{Italian Conference on Artificial Intelligence, November 30, 2021}
\end{center}
\vskip 1cm

\maketitle

\section{Introduction}
The recent rush in artificial intelligence has stimulated enormous interest
in causal inference, particularly after 
the connection of independent causal mechanisms to 
semi-supervised learning, thanks to Sch\"olkopf et al. (2012).
Historically causal inference has been formulated as a problem of
statistical testing; see Granger (1969) and Pearl (2009) for the classics.
In parallel, it has also been investigated as a physical problem. Researches
along this line include Schreiber (2000) or Palu$\rm\check s$ et al. (2001), and
Liang and Kleeman (2005). Particularly, the latter is the first one
formulated on a rigorous footing within the framework of dynamical systems, 
which yields 
an explicit solution in closed form, allowing for quantifying and normalizing
with ease the causality between dynamical events.
This short note presents a summary of this line of work, 
including the theory and several representative applications.

\section{A gentle stroll through part of the theory}
Although causality has long been studied ever since Granger's seminal 
work, its ``mathematization is a relatively recent development,'' said
Peters et al. (2017). On the other hand, Liang (2016) argued that causality
is actually 
``{\bf a real physical notion that can be rigorously derived {\it ab initio}.}''
This line of work begins with Liang and Kleeman (2005), where a discovery
about the information flow with two-dimensional deterministic systems 
was presented. A comprehensive study with generic systems
has been fulfilled recently, with explicit formulas
attained in closed forms; see Liang (2008) and Liang (2016).
These formulas have been validated with benchmark systems such as 
baker transformation, H\'enon\ map, Kaplan-Yorke map, R\"ossler system, 
to name a few.  They have also been applied to real world problems in 
the diverse disciplines such as climate science, meteorology, turbulence,
neuroscience, financial economics, etc. The following is a brief
introduction of the theory.

Consider a dynamical system, which may be either a discrete-time mapping or
a continuous-time system. For easy presentation, hereafter only the
results with the latter are shown. (The former requires the aid of the
Frobenius-Perron operator; see Liang (2016).) Let
	\begin{eqnarray}	\label{eq:stoch_gov}
	d {\ve X} = \ve F(\ve X, t)dt + \vve B(\ve X, t) d{\ve W},
	\end{eqnarray}
be a $d$-dimensional continuous-time stochastic system 
for $\ve X = (X_1, ..., X_d)$, where $\ve F = (F_1,..., F_d)$ 
may be arbitrary nonlinear functions of $\ve X$ and $t$,
${\ve W}$ is a vector of standard Wiener processes, and $\vve B  = (b_{ij})$ 
is the matrix of perturbation amplitudes 
which may also be any functions of $\ve X$ and $t$.
Assume that $\ve F$ and $\vve B$ are both differentiable with respect
to $\ve X$ and $t$. 
Now define the rate of information flow, or simply {\it information flow},
from a component $X_j$ to another component $X_i$ as the contribution of
entropy from $X_j$ per unit time in increasing the marginal entropy of
$X_i$.
We then have the following theorem (Liang, 2016): 
%
%
	\begin{thm} 
	For the system (\ref{eq:stoch_gov}),
	the rate of information flowing from $X_j$ to $X_i$ (in nats per
	unit time) is
	\begin{eqnarray}	\label{eq:Tji}
	T_{j\to i} 
        &=& -E \bracket{\frac1{\rho_i} 
		       \int_{\R^{d-2}} \Di{(F_i\rho_{\excl j})} 
				d\ve x_{\excl i \excl j}} + 
	     \frac 12 E \bracket{\frac1{\rho_i} 
	    \int_{\R^{d-2}} \DiDi {(g_{ii}\rho_{\excl j})} 
				d\ve x_{\excl i \excl j}}, \cr
	&=&
	- \int_{\R^d} \rho_{j|i} (x_j|x_i) \Di {(F_i\rho_{\excl j})} d\ve x
	  +	
	 \frac12 \int_{\R^d} \rho_{j|i} (x_j|x_i) 
		\DiDi {(g_{ii}\rho_\excl j)} d\ve x,
	\end{eqnarray}
	where $d\ve x_{\excl i \excl j}$ signifies 
	$dx_1 ... dx_{i-1} dx_{i+1} ... dx_{j-1} dx_{j+1}... dx_n$,
	$E$ stands for mathematical expectation, 
	$g_{ii} = \sum_{k=1}^n b_{ik} b_{ik}$, 
	$\rho_i = \rho_i(x_i)$ is the marginal probability density function
	(pdf) of $X_i$, $\rho_{j|i}$ is the pdf of $X_j$ conditioned on $X_i$,
	and $\rho_{\excl j} = \int_\R \rho(\ve x) dx_j$. 
	\end{thm}

Equation (\ref{eq:Tji}) has a nice property, which
forms the basis of the information flow-based causality analysis 
(Liang, 2008).
Without loss of generality, 
here the subscripts 1 and 2 are substituted by $i$ and $j$; same below.
	\begin{thm} 
		\label{thm:PNC}
	If in (\ref{eq:stoch_gov})
	neither $F_1$ nor $g_{11}$ depends on $X_2$, 
	then $T_{2\to1} = 0$.
	\end{thm}
The algorithm for the information flow-based causal inference is  
as follows: If $T_{j\to i} = 0$, then $X_j$ is not causal to $X_i$; 
otherwise it is causal, and the absolute value measures the magnitude 
of the causality from $X_j$ to $X_i$. 

Another property regards the invariance upon coordinate transformation,
indicating that the obtained information flow is an intrinsic property in
nature (Liang, 2018).
	\begin{thm}	\label{thm:invariance}
	$T_{2\to1}$ is invariant under arbitrary nonlinear transformation
of $(X_3,X_4,...,X_d)$.
	\end{thm}
As shown in Liang (2021) (and other publications), this is very important
in causal graph reconstruction. It together with Theorem~\ref{thm:PNC} 
makes it promising toward a solution of the problem of latent confounding.

For linear systems, the formula in (\ref{eq:Tji}) can be simplified.
	\begin{thm}
	In (\ref{eq:stoch_gov}), if $\ve F(\ve X) = \ve f + \vve A \ve X$,
	and the matrix $\vve B$ is constant, then
		\begin{eqnarray}
		T_{j\to i} = a_{ij} \frac {\sigma_{ij}} {\sigma_{ii}},
		\end{eqnarray}
	where $a_{ij}$ is the $(i,j)^{th}$ entry of $\vve A$ and
	$\sigma{ij}$ the population covariance between $X_i$ and $X_j$.
	\end{thm}

Notice if $X_i$ and $X_j$ are not correlated, then $\sigma_{ij}=0$, which
yields a zero causality: $T_{j\to i}=0$. But conversely it is not true.
We hence have the following corollary:
	\begin{corollary}
	In the linear sense, causation implies correlation, but not vice
	versa.
	\end{corollary}
In an explicit expression, this corollary fixes the debate on causation vs.
correlation ever since George Berkeley (1710).

In the case with only time series (no dynamical system is given), 
we have the following result (Liang, 2021): 
	\begin{thm} \label{thm:L21} 
	Given $d$ time series $X_1, X_2,...,X_d$, 
	under the assumption of a linear model with additive noise,
	the maximum likelihood estimator (mle) of (\ref{eq:Tji}) 
	for $T_{2\to1}$ is
	\begin{eqnarray}	\label{eq:T21_est}
	\hat T_{2\to1} = \frac 1 {\det\vve C} \cdot 
		       \sum_{j=1}^d \Delta_{2j} C_{j,d1}
			\cdot \frac {C_{12}} {C_{11}},
	\end{eqnarray}
where $C_{ij}$ is the sample covariance between $X_i$ and $X_j$, 
	$\Delta_{ij}$ the cofactors of the matrix $\vve C=(C_{ij})$,
	and $C_{i,dj}$ the sample covariance between $X_i$ and 
	a series derived from $X_j$ using the Euler forward differencing
	scheme:
	$\dot X_{j,n} = (X_{j,n+k} - X_{j,n}) / (k\delt)$, with $k\ge1$ 
	some integer.
	\end{thm}
Eq.~(\ref{eq:T21_est}) is rather concise in form, involving
only the common statistics, i.e., sample covariances. 
The transparent formula makes causality analysis, which otherwise 
would be complicated, very easy and computationally efficient.
Note, however, that Eq.~(\ref{eq:T21_est}) cannot replace
(\ref{eq:Tji}); it is just the maximum likelihood estimator (mle) of the latter. 
One needs to test the statistical significance before making a causal
inference based on the estimator $\hat T_{2\to1}$.

If what are given are not time series, but independent, identically
distributed (i.i.d.) panel data, it has been shown
that $\hat T_{2\to1}$ has the same form as (\ref{eq:T21_est}); see Rong and
Liang (2021).

Besides the information flow between two components, say $X_1$ and $X_2$,
it is also possible to estimate the influence of one component, say $X_1$,
on itself. Following the convention since Liang and Kleeman (2005), write
it as $dH^*_1/dt$.
	\begin{thm}
	Under a linear assumption, the mle of $dH^*_1/dt$ is
	\begin{eqnarray}
	   \widehat{\parenth{\frac{dH^*_1}{dt}}} 
	   = \frac 1 {\det\vve C} \cdot \sum_{j=1}^d \Delta_{1j} C_{j,d1}.
	\end{eqnarray}
	\end{thm}
This result, first obtained in Liang (2014), provides an efficient approach
to identifying self loops in a causal graph
(cf. Hyttinen et al., 2012). 

Statistical significance tests can be performed for the estimators. This is done
with the aid of a Fisher information matrix. 
See Liang (2014) and Liang (2021) for details.

Causality in this sense can be normalized 
in order to reveal the relative importance 
of a causal relation. See Liang (2015) for details.
Note that recently there has some similar developments along this
line, e.g., R$\phi$ysland (2012), Moij et al. (2013), 
and Mogensen et al. (2018).
We want to mention that the above formalism appears 7-13 years earlier,
and, to our best knowledge, it is the first one studying causality within
the framework of dynamical systems.


\section{Some representative applications}
The above rigorous formalism has been successfully put to application to 
many real world problems such as 
	tropical cyclone genesis prediction (Bai et al., 2018),
	 near-wall turbulence (Liang and Lozano-Dur\'an, 2016),
global climate change (Stips et al., 2016), 
	 financial analysis (Lu et al., 2020; Liang, 2015),
soil moisture-precipitation interaction (Hagan et al., 2018), 
neuroscience problems (Hristopulos et al., 2019), 
El Ni\~no (Liang et al., 2021), to name a few. 
Among these we want to particularly mention the study by Stips et al.
(2016) on CO$_2$ emission vs. global warming. 
They found that CO$_2$ emission does drive the recent global warming during
the past century, and the causal relation is one-way. However, on a time
scale of 1000 years or up, this one-way causality is 
is completely reversed, becoming a causality from air temperature to carbon dioxide. In
other words, on the paleoclimate scale, it is global warming that drives
the CO$_2$ emission! 
This remarkable result is consistent with that inferred from 
the recent ice-core data from Antarctica.

Another interesting application (Liang, 2015) regards the
relation between the two corporations IBM and GE, using the time series of
US stocks downloaded from ${YAHOO!} \atop {\rm {finance}}$. 
Overall the causality between the two is insignificant, but
if a running time causality analysis is performed, there appears a strong, 
almost one-way causality from IBM to GE 
in 70's, starting from 1971. This abrupt one-way causality out of blue
reveals to us an old story about ``Seven Dwarfs and a Giant'': 
In 50-60's, GE was believed to be the biggest computer user outside the U.S.
Federal Government; to avoid relying on IBM the computer ``Giant'', 
it together with six other companies (``Seven Dwarfs'') 
began to build mainframes. GE itself of course is a giant, but in the computer
market, it is just a ``Dwarf''. Since it could not beat IBM, in 1970, it sold 
its computer division. 
As a result, starting from 1971, it had to rely on IBM again. 
This is the story behind the jump in $T_{IBM\to GE}$ from 1970 to 1971.
While the story has almost gone to oblivion,
this finding, which is solely based on a simple causality analysis 
of two time series with Eq.~(\ref{eq:T21_est}), is really remarkable.

	\begin{figure}[h]
	\begin{center}
	\includegraphics[angle=0, width=0.75\textwidth]{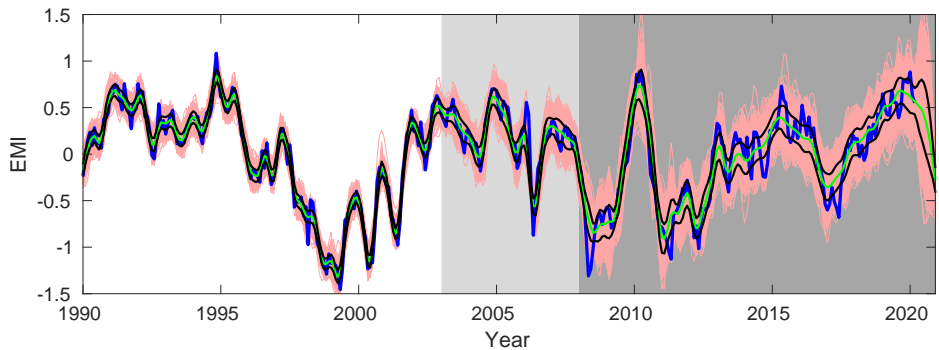}
	\caption{El Ni\~no prediction has become a benchmark problem for
	the testing of machine learning algorithms. The present wisdom is 
	that El Ni\~no may be predicted at a lead time of 1-2 years.
	Shown here are 1000 predictions (pink) of the El Ni\~no Modoki
	index (EMI) as described in the text. 
	Overlaid are the observed EMI (blue), the mean of the
	realizations (cyan). The light shading marks the period for
	validation, while the darker shading marks the prediction period.
	(From Liang et al., 2021.)}
	\end{center}
	\end{figure}

The latest application on the prediction of El Ni\~no Modoki, or
Central-Pacific type El Ni\~no, is also a remarkable one.
El Ni\~no is a climate mode that has been linked to many hazards globewide,
e.g., flooding, drought, wild fires, heat waves, etc.
Its accurate forecasting is of great importance to many sectors of our
society such as agriculture, energy, hydrology, to name several. With its
societal importance and the elegant setting, El Ni\~no prediction has become 
a testbed for AI algorithms.

Currently the widsom for El Ni\~no prediction is that it may be 
predictable at a lead time of 1-2 years.
However, there still exists much uncertainty;
an example is the 2014-16 ``Monster El Ni\~no,'' almost all projections
fell off the mark. Among the El Ni\~no varieties, it is 
believed that El Ni\~no Modoki is particularly difficult to predict.

A striking breakthrough has just been made. 
Liang et al. (2021) took advantage of the quantitative nature of
the above information flow-based causality analysis, 
and identified a delayed causal pattern, i.e., the
structure of the information flow from the
solar activity to the sea surface temperature, very similar to the El
Ni\~no Modoki mode. They then conjectured that, based on the series of
sunspot numbers, El Ni\~no Modoki should be predictable. This is
indeed the case, and, remarkably, the prediction can be at a lead time of
as long as 10 years or up! 
This remarkable progress, among others, is a result of the rigorously
formulated quantitative causality analysis.

{

}


\begin{thebibliography}{}

\bibitem{Bai2017} Bai, C., R. Zhang, and Coauthors (2018). Forecasting the
tropical cyclone genesis over the Northwest Pacific through identifying the
causal factors in cyclone-climate interactions. {\em J. Atmos. Ocean.
Tech.}, 35, 247-259.

 \bibitem{Berkeley} Berkeley, G. {\em A Treatise Concerning the Principles
of Human Knowledge}; Aaron Rhames: Dublin, Ireland, 1710.


\bibitem{Eberhardt} Hyttinen, A., F. Eberhardt, P.O. Hoyer (2012). 
Learning linear cyclic causal models with latent variables.
{\em Journal of Machine Learning Research}, 13, 3387-3439.

\bibitem{Granger1969}
Granger, C.W.J. (1969). Investigating causal relations by econometric models
and cross-spectral methods. {\em Econometrica}, \emph{37}, 424--438.

\bibitem{Hagan2019}
Hagan, D.F.T.; Wang, G.; Liang, X.S.; Dolman, H.A.J. (2019) A time-varying
causality formalism based on the Liang-Kleeman information flow for
analyzing directed interactions in nonstationary climate systems.
{\em J. Clim.}, \emph{32}, 7521--7537.


\bibitem{Hristopulos2019}
Hristopulos, D.T.; Babul, A.; Babul, S.; Brucar, L.R.; Virji-Babul, N.
(2019)
 Disrupted information flow in resting-state in adolescents with 
sports related concussion. {\em Front. Hum. Neurosci.}, 
\emph{13}, 419.  doi:10.3389/fnhum.2019.00419.

\bibitem{Scholkopf} Janzing, D., B. Sch\"olkopf (2017). Detecting
confounding in multivariate linear models via spectral analysis.
{\em J. Causal Inference}, 6(1), DOI:10.1515/jci-2017-0013.


\bibitem{LK05} Liang, X.S., R. Kleeman (2005).
Information transfer between dynamical system components.
{\rm Phys. Rev. Lett.}, 95(24), 244101.

\bibitem{Liang2008} Liang, X.S. Information flow within stochastic
dynamical systems. {\em Phys. Rev. E} \textbf{2008}, \emph{78}, 031113.


\bibitem{Liang2014} Liang, X.S. (2014). Unraveling the cause-effect relation
between time series. {\em Phys. Rev. E}, 90, 052150.

\bibitem{Liang2015} Liang, X.S. (2015). Normalizing the causality between time
series. {\em Phys. Rev. E} \textbf{2015}, \emph{92}, 022126.


\bibitem{Liang2016} Liang, X.S. (2016). Information flow and causality as
rigorous notions {ab initio}. {\em Phys. Rev. E}, 94, 052201.

\bibitem{stanford2016} Liang, X.S., A. Lozano-Dur\'an (2016). A preliminary
study of the causal structure in fully developed near-wall turbulence. 
{\it Proceedings of the Summer Program 2016}, 233-242. Center for Turbulence
Research, Stanford University, CA, USA. 

\bibitem{Liang2018} Liang, X.S. (2018).  Information flow with respect to
relative entropy. {\em Chaos}, \emph{28}, 075311.

\bibitem{Liang2019} Liang, X.S. (2019). A study of the cross-scale
causation and information flow in a stormy model mid-latitude atmosphere.
{\it Entropy}, 21, 149.

\bibitem{Liang2021} Liang, X.S. (2021). Normalized multivariate time series
causality analysis and causal graph reconstruction. {\em Entropy}, 23, 679.

\bibitem{Liang_enso2021} Liang, X.S., F. Xu, Y. Rong, R. Zhang, X. Tang,
and F. Zhang (2021). El Ni\~no Modoki can be mostly predicted more than 10
years ahead of time. {\em Sci. Rep.}, 11, 17860.

\bibitem{Lu2021}
Lu, X., K. Liu, et al. (2020). The break point-dependent causality between the
cryptocurrency and emerging stock markets. {\em Economic Computation and
Economic Cybernetics Studies and Research}, 54(4), 203-216.

\bibitem{Mooij2013}
Mooij, J.M.; Janzing, D.; Heskes, T.; Sch\"olkopf, B. (2013)   From ordinary
 differential equations to structural causal models: The deterministic
 case. In  Proceedings of the 29th Annual Conference on Uncertainty in
 Artificial Intelligence, {Bellevue, WA, USA, 11--15, July}, 
 pp. 440--448.
 
 \bibitem{Mogensen2018}
Mogensen, S.W.; Malinksky, D.; Hansen, N.R. (2018) Causal learning for
partially observed stochastic dynamical systems. In Proceedings of
the 34th Conference on Uncertainty in Artificial Intelligence (UAI), {Monterey, CA, USA, 6--10 August} 2018.

\bibitem{Palus2001} Palu$\rm\check s$, M.; Kom\'arek, V.;
Hrn$\rm\check c$i$\rm\check r$, Z.; $\rm\check S$t$\rm\check e$rbov\'a, K.
(2001)
Synchronization as adjustment of information rates: Detection from
bivariate time series. {\em Phys. Rev. E}, \emph{63}, 046211.

\bibitem{Pearl2009} Pearl, J. {\em Causality: Models, Reasoning,
and Inference}, 2nd ed; Cambridge University Press: New York, NY, USA, 2009.

\bibitem{Pethel2011}
Hahs, D.W.; Pethel, S.D. (2011) Distinguishing anticipation from
causality: Anticipatory bias in the estimation of information flow.
{\em Phys. Rev. Lett.}, \emph{107}, 12870.


\bibitem{Peters2017}
Peters, J., Janzing, D., Sch\"olkopf, B. (2017) {\it Elements of Causal
Inference: Foundations and Learning Algorithms}. The MIT Press, Cambridge,
MA, USA.


\bibitem{Rong2021} Rong, Y., X.S. Liang (2021). Panel data causal inference
using a rigorous information flow analysis for homogeneous, independent
and identically distributed datasets. {\em IEEE Access}, 9, 47266-47274.


\bibitem{Reichenbach}
Reichenbach, H. (1956). {\em The direction of time}. University of
California Press, Berkeley.

\bibitem{Roysland2012}
R$\crossout o$ysland, K. (2012). Counterfactual analyses with graphical
models based on local independence. {\em The Annals of Statistics},
40(4):2162-2194.

\bibitem{Scholkopf2012} Sch\"olkopf, B.; Janzing, D.; Peters, J.;
Sgouritsa, E.; Zhang, K.; Mooij, J.M.
On causal and anticausal learning. In {Proceedings of the 29th
International Conference on Machine Learning (ICML)}, 
Edinburgh, Scotland, June 26-July 1, 2012; pp.1255-1262.

\bibitem{Schreiber2000}
Schreiber, T.  Measuring information transfer. {\em Phys. Rev. Lett.} \textbf{2000}, 
\emph{85}, 461.


\bibitem{Spirtes2016}
Spirtes, P., K. Zhang (2016). Causal discovery and inference: concepts and recent
methodological advances. {\em Appl. Informatics},
\emph{3}, 3.


\bibitem{Stips2016} Stips, A.; Macias, D.; Coughlan, C.; Garcia-Gorriz, E.; Liang, X.S.   On the causal structure between CO$_2$ and global
temperature. {\em Sci. Rep.} \textbf{2016}, \emph{6}, 21691.




%

\end{thebibliography}
\end{document}